\documentclass[aps,prl,twocolumn,superscriptaddress,nofootinbib,floatfix]{revtex4}
\usepackage{graphics}
\usepackage{amsmath}
\usepackage{epsfig}
\usepackage{helvet}
\usepackage{amssymb}

\newcommand{\be}{\begin{equation}}
\newcommand{\ee}{\end{equation}}
\newcommand{\bea}{\begin{eqnarray}}
\newcommand{\eea}{\end{eqnarray}}

\newcommand{\sSO}{{\mathcal{Q}}\hspace{-.3mm}_{^\sharp}\hspace{-.1mm}}
\newcommand{\SO}{\mathcal{Q}}
\newcommand{\sso}{\hspace{-.2mm}_{^\sharp}\hspace{-.4mm}}
\newcommand{\str}{\mbox{tr}\hspace{-.3mm}_{^\sharp}\hspace{-.3mm}}

\newcommand{\tr}{\mbox{tr}}
\newcommand{\bra}[1]{\mbox{$\langle #1 |$}}
\newcommand{\ket}[1]{\mbox{$| #1 \rangle$}}

\newcommand{\proj}[1]{\mbox{$|#1\rangle \!\langle #1 |$}}

\newcommand{\sbra}[1]{\mbox{${_{_\sharp}}\hspace{-.8mm} \langle #1 |$}}
\newcommand{\sket}[1]{\mbox{$| #1 \rangle\hspace{-.5mm}_{_\sharp}$}}
\newcommand{\sbraket}[2]{\mbox{$_{_\sharp}\hspace{-.8mm}\langle #1  | #2 \rangle\hspace{-.5mm}_{_\sharp}$}}
\newcommand{\sproj}[1]{\mbox{$|#1\rangle \hspace{-.5mm}_{_\sharp}\hspace{-.8mm}\langle #1 |$}}
\def\ch{\raisebox{0.3ex}{\mbox{$\chi$}}}

\begin{document}
\title{
Mixed-state dynamics in one-dimensional quantum lattice systems:\\ a time-dependent superoperator renormalization algorithm.
}

\author{Michael Zwolak} 
\email{zwolak@caltech.edu}
\affiliation{Institute for Quantum Information, California Institute of Technology, Pasadena, California 91125}
\author{Guifr\'e Vidal} 
\email{vidal@iqi.caltech.edu}
\affiliation{Institute for Quantum Information, California Institute of Technology, Pasadena, California 91125}
\date{\today}
\begin{abstract}
We present an algorithm to study mixed-state dynamics in one-dimensional quantum lattice systems. The algorithm can be used, e.g., to construct thermal states or to simulate real time evolution given by a generic master equation. Its two main ingredients are ($i$) a {\em superoperator} renormalization scheme to efficiently describe the state of the system and ($ii$) the time evolving block decimation technique to efficiently update the state during a time evolution. The computational cost of a simulation increases significantly with the amount of correlations between subsystems but it otherwise depends only linearly on the system size. We present simulations involving quantum spins and fermions in one spatial dimension.

\end{abstract}

\maketitle

The most interesting quantum phenomena involve strongly correlated many-body systems, but studying such systems --- a central task in the areas of condensed matter physics, quantum field theory and, since recent years, also quantum information science \cite{book,Preskill}--- has too often proven a formidable challenge. Indeed, in quantum many-body theory only a few exact solutions are available, while most analytical approximations remain uncontrolled. As a consequence, numerical calculations are of great importance. But even these suffer from a severe computational obstacle: an exponential growth of degrees of freedom with the system size that renders the direct simulation of most quantum systems prohibitively inefficient.

And yet, ingenious methods such as quantum Monte Carlo techniques \cite{suzuki} can be used to approximately evaluate, e.g., certain ground state properties in quantum lattice models. In one dimensional lattices, strikingly accurate results for quantities such as ground state energies and two-point correlators can be obtained by using White's {\em density matrix renormalization group} (DMRG) \cite{dmrg}, a technique that has dominated most numerical research in the field since its invention more than a decade ago. Generalizations of the DMRG have also yielded accurate low energy spectra \cite{dmrg2} or allowed for the simulation of real time evolution for small times \cite{dmrg3}. 

Recently, the {\em time evolving block decimation} (TEBD) algorithm \cite{vid1} has been proposed to simulate real time evolution in one-dimensional quantum lattice systems. This technique can be easily adapted to standard DMRG implementations \cite{white,daley} and seems to be very efficient \cite{white,daley,jaksch}. As in DMRG, a decisive factor in the performance of the TEBD method is that not a lot of entanglement is present in the system, a condition that is ordinarily met in one-dimensional lattices at low energies \cite{vid1}.

In this paper we extend the TEBD algorithm to handle mixed states. We describe how to efficiently simulate, in one-dimensional quantum lattice systems, real time Markovian dynamics as given by a (possibly time-dependent) master equation made of arbitrary nearest neighbor couplings. By considering evolution in imaginary time, the present extension can also be used to construct thermal states for any given temperature. Thus, we show how to numerically explore non-equilibrium many-body dynamics under realistic conditions, including the effects of finite-temperature and decoherence.

A key observation for the success of the algorithm is that in one spatial dimension many states of interest, including thermal states and local perturbations thereof, contain only a restricted amount of correlations between subsystems, in a sense to be further specified. This fact, parallel to the restricted amount of entanglement observed in the pure-state case, allows us to introduce an efficient decomposition for the state of the system, referred to as {\em matrix product decomposition} (MPD). The MPD is nothing but a mixed-state version of a matrix product state \cite{mps}, and as such, we can use the TEBD to update it during a time evolution. It also follows that our scheme can again be fully incorporated into standard DMRG implementations without much programming effort \cite{white,daley}.

We consider a generic one dimensional quantum lattice made of $n$ sites, labeled by index $l$, $l \in \{1,\cdots,n\}$, each one described by a local Hilbert space $\mathbb{H}^{[l]} \cong \mathbb{C}_d$ of finite dimension $d$. We assume the evolution of the $n$ sites, in a global state $\rho$, is given by a master equation \cite{Preskill}
\bea
\dot{\rho}
&=& \mathcal{L}[\rho] \label{eq:master}\\
&=& -i[H,\rho] + \sum_{\mu} \left(L_{\mu}\rho L_{\mu}^{\dagger} - \frac{1}{2} L_{\mu} L_{\mu}^{\dagger}\rho - \frac{1}{2} \rho L_{\mu} L_{\mu}^{\dagger}\right), \nonumber
\eea
where $H$ and $L_{\mu}$ are the Hamiltonian and Lindblad operators, and where we require that the (possibly time-dependent) Lindbladian superoperator $\mathcal{L}$ further decompose into terms involving at most two contiguous sites,
\be
{\cal L}[\rho] = \sum_l {\cal L}_{l,l+1}[\rho].
\label{eq:nearest}
\ee

{\em Reduced superoperators.}--- A pure state evolution is described by a vector $\ket{\Psi}$ in the $n$-fold tensor product of $\mathbb{C}_d$. Let us divide the $n$ sites into two blocks, denoted $L$ (left) and $R$ (right). Then DMRG and TEBD consider reduced density matrices, e.g., that of block $L$,
\be
\ket{\Psi} \in \mathbb{C}_d^{~\otimes n}  ~~~\longrightarrow ~~~  \rho^{[L]} \equiv \mbox{tr}_R (\proj{\Psi}) ~ \in \mathbb{L}(\mathbb{H}^{[L]}),
\label{eq:nosuper}
\ee
where $\mathbb{L}(\mathbb{H})$ denotes the set of linear mappings on $\mathbb{H}$ or, equivalently, the complex vector space of dim$(\mathbb{H})\times$dim$(\mathbb{H})$ matrices. Here we are concerned with the evolution of a mixed state, which requires more notation. For each site $l$, let $\mathbb{K}^{[l]} \cong \mathbb{L}(\mathbb{H}^{[l]})\cong \mathbb{C}_{d^2}$ denote the vector space of $d\times d$ complex matrices. We switch into representing a density matrix $\sigma\in \mathbb{L}(\mathbb{H})$ as a ``superket'' $\sket{\sigma}\in \mathbb{K}\cong \mathbb{L}(\mathbb{H})$, while a superoperator $\SO \in \mathbb{L}(\mathbb{L}(\mathbb{H}))$ is regarded as a linear mapping $\sSO \in \mathbb{L}(\mathbb{K})$ \cite{notation},
\be
\left. \begin{array}{llll}
\ket{\Phi} ~ &\in& ~ \mathbb{H}\\
~\sigma ~ &\in& ~\mathbb{L}(\mathbb{H}) \\
~\SO ~ &\in& ~\mathbb{L}(\mathbb{L}(\mathbb{H}))
\end{array}
\right\}\rightarrow \left\{
\begin{array}{llll} 
\sket{\Phi} ~&\in& ~\mathbb{K}  \\
\sket{\sigma} ~&\in& ~\mathbb{K} \\
~\sSO ~&\in& ~\mathbb{L}(\mathbb{K}), 
\end{array}
\right.
\ee
where $\sket{\Phi}\equiv \sket{\proj{\Phi}}$. 
For $d\times d$ matrices $A$ and $B$, 
the scalar product $\sbraket{}{}$ 
between superkets $\sket{A}$ and $\sket{B}$, and the action of $\sSO$ on $\sket{A}$, are defined through
\be
\sbraket{A}{B} \equiv \frac{1}{d}\tr(A^\dagger B),~~~~~  
\sSO\sket{A} \equiv \sket{\SO[A]}.
\ee
Also, if $\SO$ is a superoperator on a bipartite space $\mathbb{H}^{[L]}\otimes \mathbb{H}^{[R]}$ and $\{\sket{M_\mu}\}$ is an orthonormal basis in $\mathbb{K}^{[R]} \cong \mathbb{L}(\mathbb{H}^{[R]})$, we define the partial trace of $\sSO$ over block $R$ as 
\be
\str_R (\sSO) \equiv \sum_{\mu} \sbra{M_\mu}\sSO\sket{M_\mu}.
\ee
Finally, let $\rho \in \mathbb{L}(\mathbb{C}_d^{~\otimes n})$ be the state of the $n$-site lattice and $\sket{\rho}$ its superket. We define the {\em reduced superoperator} for a block of sites, say for block $L$, as (see example \cite{example})
\be
\sket{\rho} \in (\mathbb{C}_{d^2})^{\otimes n} ~\longrightarrow ~ \sSO^{[L]}\equiv  \str_R (\sproj{\rho})~ \in \mathbb{L}(\mathbb{K}^{[L]}),
\label{eq:super}
\ee 
in analogy with (\ref{eq:nosuper}), and rewrite 
equation (\ref{eq:master}) as
\be
\ket{\dot{\rho}}\hspace{-.5mm}_{_\sharp}
= \mathcal{L}\sso \sket{\rho},
\label{eq:master2}
\ee
which parallels the Schr\" odinger equation 
$ \ket{\dot{\Psi}}
= -i H\ket{\Psi}$.

{\em Renormalization of reduced superoperators.}--- Given blocks $L$ and $R$, the Schmidt decomposition of $\sket{\rho}$ reads
\be
\sket{\rho} = \sum_{\alpha=1}^{\ch\sso} \lambda\sso_{\alpha}\sket{M^{[L]}_\alpha}\otimes\sket{M^{[R]}_\alpha},  ~~~ \lambda\sso_{\alpha} \geq\lambda\sso_{\alpha+1} \geq 0,
\label{eq:Schmidt}
\ee
where the Schmidt superkets $\{\sket{M^{[L,R]}_\alpha}\}$ fulfill
\bea
\sSO^{[L]}\sket{M^{[L]}_\alpha} = (\lambda\sso_{\alpha})^2 \sket{M^{[L]}_\alpha}, ~~\sSO^{[L]}\equiv  \str_R (\sproj{\rho}), \\
\sSO^{[R]}\sket{M^{[R]}_\alpha} = (\lambda\sso_{\alpha})^2 \sket{M^{[R]}_\alpha},~~\sSO^{[R]}\equiv  \str_L (\sproj{\rho}).
\eea
The rank $\ch\sso$ of the reduced superoperators $\sSO^{[L]}$ and $\sSO^{[R]}$ measures the amount of correlations between blocks $L$ and $R$ \cite{preparation}. In principle its value is only bounded above by the dimensions of $\mathbb{K}^{[L]}$ and $\mathbb{K}^{[R]}$, which grow exponentially in the number of sites. However, as the examples below illustrate, many situations of interest involving one-dimensional mixed-state dynamics are only {\em slightly correlated}, in that the coefficients $\{\lambda\sso_{\alpha}\}$ decay very fast with $\alpha$. That is, a good approximation to $\sket{\rho}$ can be obtained by truncating (\ref{eq:Schmidt}) so that only a relatively small number of terms are considered.

Thus, whereas an efficient description of $\ket{\Psi}$ in (\ref{eq:nosuper}) is achieved both in DMRG and TEBD by conveniently decimating, say, the block space $\mathbb{H}^{[L]}$ supporting the reduced density matrix $\rho^{[L]}$, our efficient description of $\rho$ is based instead on decimating the block space $\mathbb{K}^{[L]} \cong \mathbb{L}(\mathbb{H}^{[L]})$ supporting the reduced superoperator $\SO^{[L]}$ in (\ref{eq:super}) \cite{intuition,compare}. 


\begin{figure}
\epsfig{file=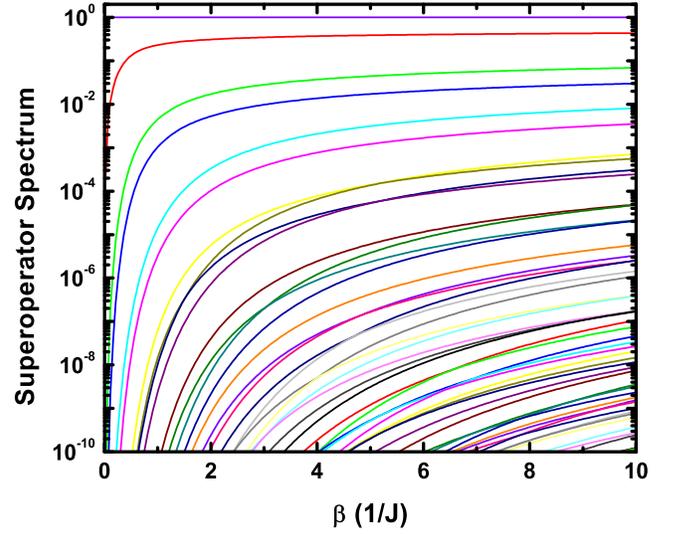,width=8.5cm}
\caption{Quantum Ising chain with transverse magnetic field, Eq. (\ref{eq:Ising}), at finite temperature, with local dimension $d=2$, $n=100$ sites and effective $\chi\sso = 80$. At zero temperature this model corresponds to a quantum critical point. The spectrum $\{ \lambda{\sso}_{\alpha}^{2} \}$ of the reduced superoperator $\sSO^{[L]}$ for the left $n/2=50$ spins is plotted as a function of $\beta \in [0,10/J]$ (only the largest $52$ $\lambda\sso_{\alpha}$'s are shown). For any inverse temperature $\beta$, a fast decay of the eigenvalues $\lambda\sso_{\alpha}^2$ in $\alpha$ ensures that the state can be accurately approximated by a MPD with small effective $\chi{\sso}$.
}
\label{fig:thermal}
\end{figure}


{\em Matrix Product Decomposition and TEBD.}--- 
We regard the $n$-site $\rho$ as a vector $\sket{\rho}$ in the $n$-fold tensor product of $\mathbb{C}_{d^2}$, while the master equation (\ref{eq:master2}) is formally identical to the Schr\" odinger equation. Simulating mixed-state dynamics can therefore be achieved by conveniently adapting the pure-state techniques of \cite{vid1}. More specifically, given an orthonormal basis $\{\sket{i_l}\}$ of $\mathbb{K}^{[l]}$ for each site $l$ ($l=1, \cdots, n$), we expand $\sket{\rho}$ as \cite{real}
\be
\sket{\rho} = \sum_{i_1=0}^{d^2-1}\!\cdots \!\sum_{i_n=0}^{d^2-1} c_{i_1 \cdots i_n} ~\sket{i_1}\otimes \cdots \otimes \sket{i_n}.
\label{eq:local}
\ee
We choose $\sket{0_l} = \sket{I/d}$ to be proportional to the identity (that is, as a mapping on $\mathbb{H}^{[l]}$), so that physical normalization of $\rho$, $\tr(\rho) = 1$, corresponds to $c_{0\cdots 0} = 1$. We use a MPD for the coefficients $c_{i_1 \cdots i_n}$,
\bea
c_{i_1i_2 \cdots i_n} = \!\!\! \sum_{\alpha_1,\cdots,\alpha_{n\!-\!1}}\!\!\! \Gamma_{\alpha_1}^{[1]i_1}  \lambda{\sso}^{[1]}_{\alpha_1} \Gamma_{\alpha_1 \alpha_2}^{[2]i_2}  \lambda{\sso}^{[2]}_{\alpha_2} \Gamma_{\alpha_2 \alpha_3}^{[3]i_3}  \cdots  \Gamma_{\alpha_{n\!-\!1}}^{[n]i_n},
\label{eq:superdeco}
\eea
which can be built through a succession of Schmidt decompositions of $\sket{\rho}$ (see \cite{vid1} for details). Finally, we can use the TEBD method to update the tensors $\{\Gamma^{[l]}\}$ and $\{\lambda{\sso}^{[l]}\}$ during an evolution of the form (\ref{eq:master})-(\ref{eq:nearest}) \cite{nonunitary}.

{\em Example 1: Thermal state.---} Given a nearest neighbor Hamiltonian $H$ and an inverse temperature $\beta \equiv 1/kT$, a mixed state of interest is the thermal state
\be
\rho_{\beta} \equiv \frac{e^{-\beta H}}{Z(\beta)} = \frac{1}{Z(\beta)}\sum_{s}e^{-\beta E_s}\proj{E_s},
\ee
where $Z(\beta) \equiv tr(e^{-\beta H})$ is the partition function. One can numerically simulate $\rho_\beta$ by attempting to compute all relevant energy eigenstates $\ket{E_s}$ and averaging them with weights $e^{-\beta E_s}/Z(\beta)$. A very simple and efficient alternative is to build a MPD for $\sket{\rho_\beta}$ by simulating an imaginary time evolution from the completely mixed state,
\be
\sket{e^{-\beta H}} = \exp({-\beta \mathcal{T}_{\sso}})\sket{I},
\ee
where superket $\sket{I}$ and superoperator $\mathcal{T}\sso$ correspond to
\be
\sket{I} = \sket{I_1}\otimes\cdots \otimes \sket{I_n}, ~~\mathcal{T}[A] \equiv \frac{1}{2}(HA + AH).
\ee
Indeed, $\exp(-\beta \mathcal{T}_{\sso})$ can be Trotter expanded into transformations involving only two adjacent sites, and the MPD can be therefore updated using the TEBD  \cite{advise}. Notice that a single run of the simulation builds the thermal state $\rho_{\beta'}$ for any intermediate value of $\beta' \in [0,\beta]$. Fig. (\ref{fig:thermal}) corresponds to thermal states for a quantum Ising model with transverse magnetic field, 
\be
H = \sum_{l=1}^{n-1} \sigma^x_l\otimes \sigma^x_{l\!+\!1} + \sum_{l=1}^{n} \sigma^z_{l}.
\label{eq:Ising}
\ee


\begin{figure}
\epsfig{file=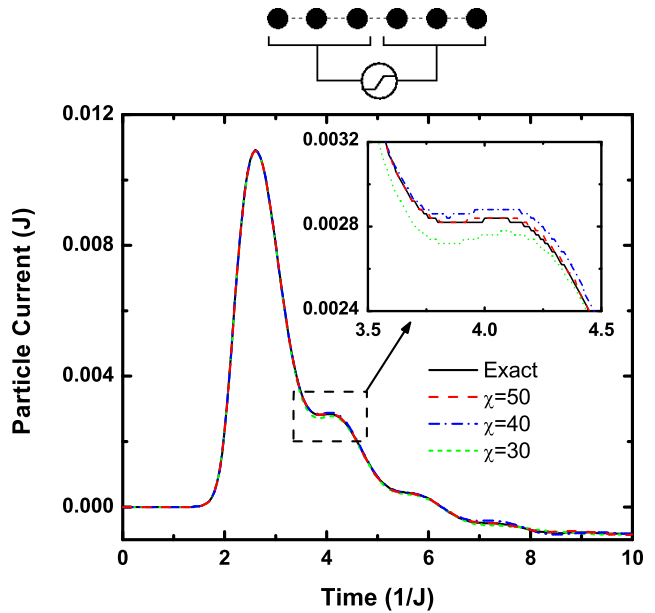,width=8.5cm}
\caption{Fermionic lattice of Eq. (\ref{eq:masterfermions}) at finite temperature $\beta = 1/J$, dephasing $\gamma = 0.4J$, and with $d=2$, $n=100$. The particle current, see Eq. (\ref{eq:current}), is due to a time-dependent applied bias $\mu(t)$ with turn-on time $t_0 = 2/J$ and rise time $t_s = 0.1/J$. Simulations with an effective $\chi\sso = 30,40,50$ show rapid convergence to the exact solution. This convergence will be addressed in more detail in \cite{preparation}.}
\label{fig:fermion1}
\end{figure}


\begin{figure}
\epsfig{file=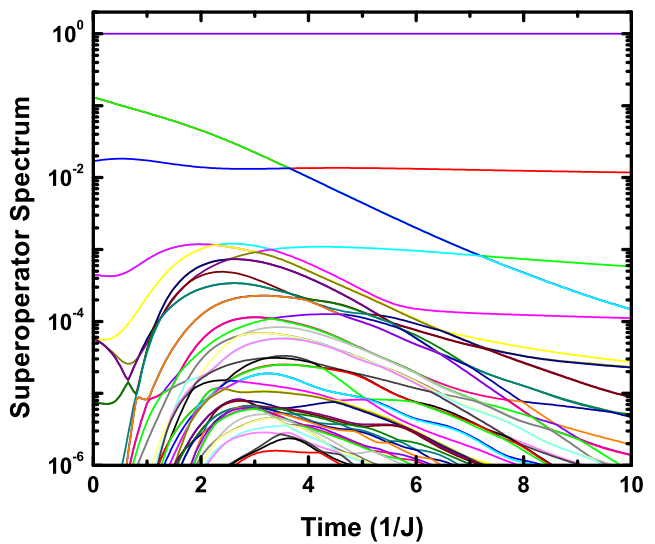,width=8.5cm}
\caption{Same system as in Fig. (\ref{fig:fermion1}). The spectrum $\{\lambda\sso_{\alpha}^2\}$ of the reduced superoperator $\sSO^{[L]}$ for the left $n/2=50$ sites is plotted as a function of time. The number of relevant eigenvalues $\lambda\sso_{\alpha}^2$ (say above $10^{-6}$) increases as the applied bias is turned on, but remains small throughout the evolution and it even decreases for long times.}
\label{fig:fermion2}
\end{figure}


\begin{figure}
\epsfig{file=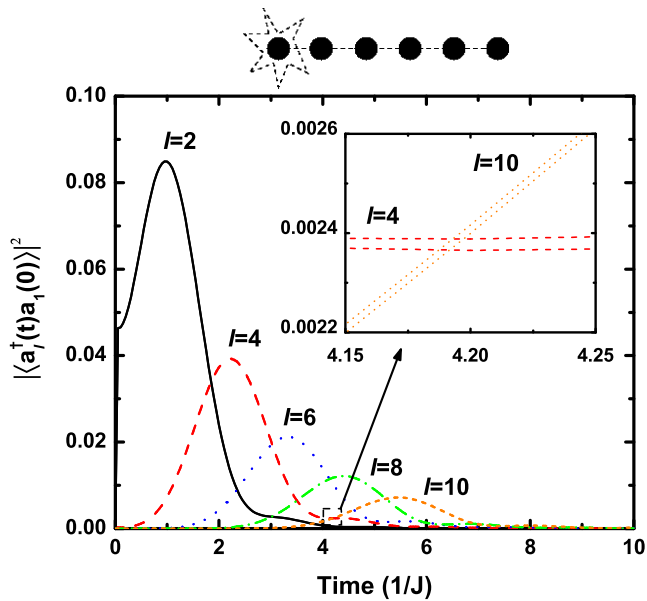,width=8.5cm}
\caption{Fermionic lattice of Eq. (\ref{eq:masterfermions}) at finite temperature $\beta = 1/J$, dephasing $\gamma = 0.4J$, and with $d=2$, $n=100$ and no applied bias, $\mu(t)=0$. Unequal time, two-point correlator (\ref{eq:unequal}) for $l=2,4,6,8, 10$ and $t\in [0,10/J]$. The results corresponding to an effective $\chi\sso = 40$ and $50$ practically overlap at all times, as the inset shows.}
\label{fig:unequal}
\end{figure}


{\em Example 2: Time-dependent master equation.---} 
We consider a lattice of $n=100$ sites loaded with $n/2$ fermions that evolve according to a Lindbladian 
\be
\mathcal{L}[\rho] = -i[H,\rho] + \gamma \sum_{l=1}^{n} (n_l\rho n_l -\frac{1}{2}\rho n_l^2 -\frac{1}{2}  n_l^2\rho), 
\label{eq:masterfermions}
\ee
$n_l \equiv a_l^{\dagger}a_l$, where the last term accounts for phase damping and the Hamiltonian part corresponds to hoping between adjacent sites and a time-dependent on-site energy,
\be
H = -J\sum_{l=1}^{n-1} (a_l^\dagger a_{l+1} + h.c.) - \mu(t)(\sum_{l=1}^{n/2} n_l - \!\!\!\!\sum_{l=n/2+1}^{n} \!\!\!\!n_l),
\ee
where $\mu(t) \equiv \mu_0[e^{-\frac{t-t_0} {t_s} } +1]^{-1}$ introduces a bias $\mu_0$ between the left and right halves of the lattice at $t = t_0$.
Fig. (\ref{fig:fermion1}) shows the particle current 
\be
-2 \, \mbox{Im } \langle a^{\dagger}_{50}(t)a_{51}(t) \rangle
\label{eq:current}
\ee
from the right half of the lattice to the left half as a result of switching on the bias. The numerical results have been obtained by mapping the fermionic lattice into a spin lattice through a Jordan-Wigner transformation and by simulating the resulting spin lattice model, which still contains only nearest neighbor couplings, using the TEBD algorithm. The exact solution can be efficiently obtained by numerically integrating the time evolution of two point-correlators. Comparison with the exact solution shows that the simulations for small effective $\ch\sso = 30,40,50$ are remarkably accurate and converge for increasing value of $\ch\sso$. The fast decay of the spectrum in Fig. (\ref{fig:fermion2}) justifies this convergence.


{\em Example 3: Unequal-time correlators.---} For the above fermion system with no bias, $\mu(t) = 0$, and finite temperature, we finally consider the expectation value
\be
\langle {a_{l}^\dagger(t)} a_{1}(0)\rangle = \tr\left({a_{l}^\dagger} \mathcal{E}_{t}[a_1 \rho]\right),
\label{eq:unequal}
\ee
where $\mathcal{E}_t$ is the time evolution operator resulting from the master equation. The simulation (see Fig. (\ref{fig:unequal})) is achieved as follows: 
$(i)$ the initial state of the system, a thermal state with $\beta = 1/J$, is obtained by evolution in imaginary time as explained in Example 1;
($ii$) the annihilation operator $a_1$ is applied to the initial state $\rho$ to obtain $a_1\rho$; 
($iii$) $a_1\rho$ is evolved in time according to $\mathcal{E}_t$; 
($iv$) the creation operator $a_l^{\dagger}$ is applied on $\mathcal{E}_t[a_1\rho]$; and 
($v$) the trace of the resulting operator $a_l^{\dagger}\mathcal{E}_t[a_1\rho]$ is computed. Each of these steps can be performed efficiently by using a MPD and the update techniques of \cite{vid1}. In this particular case the Lindbladian $\mathcal{L}$ is time-independent and we can integrate Eq. (\ref{eq:master2}), so that step ($iii$) becomes
\be
\sket{\mathcal{E}_t[a_1\rho]} = \mathcal{E}_{t\sso} \sket{a_1\rho} =  \exp({\mathcal{L}_{\sso}t}) \sket{a_1\rho}.
\ee
Because of property (\ref{eq:nearest}), $\exp({\mathcal{L}_{\sso} t})$ can be Trotter-expanded into small transformations involving only two adjacent sites, and therefore it can be implemented using the TEBD.

We have presented an extension of the TEBD algorithm to mixed states. With specific examples involving spins and non-interacting fermions, we have shown how to ($i$) construct thermal states; ($ii$) evolve a state in time according to a time-dependent master equation; and  $(iii)$ compute unequal time correlation functions. The algorithm can be used for generic one-dimensional lattice systems, including interacting fermions and bosons as explored in \cite{preparation}.

MZ acknowledges support from an NSF Graduate Fellowship. The authors acknowledge support from U.S. NSF grant no. EIA-0086038.

See also F. Verstreate {\em et al.}, cond-mat/0406426.

{\em Note:} The extension of the TEBD method presented here was outlined in [G. Vidal, quant-ph/0301063 v2 (2003)] but excluded from the first reference in \cite{vid1} by request of a referee.


\begin{thebibliography}{99}
\bibitem{notation} We use the subscript $\sharp$ ``sharp'' to denote operators [superoperators] when represented as superkets [respectively mappings between superkets].
\bibitem{example} As an example, let us consider a mixed state of two two-level systems,
\be
\rho = \sum_{i_1,i_2 = 0}^{3} c_{i_1i_2} \sigma_{i_1}\otimes\sigma_{i_2} ~~~\in ~~\mathbb{L}(\mathbb{C}_2\otimes\mathbb{C}_2)
\ee
where $\sigma_i = \{I, \sigma_x,\sigma_y,\sigma_z\}$ are the identity and Pauli matrices. The corresponding superket reads
\be
\sket{\rho} = \sum_{i_1,i_2 = 0}^{3} c_{i_1i_2} \sket{\sigma_{i_1}}\otimes\sket{\sigma_{i_2}}~~~ \in~ \mathbb{C}_4\otimes\mathbb{C}_4,
\ee
where $\sbraket{\sigma_i}{\sigma_j} = \delta_{ij}$, whereas the reduced superoperator for the first system is
\bea
\sSO &=& \str_2 \sproj{\rho}\\ 
&=& \sum_{i_2} \sbraket{\sigma_{i_2}}{\rho} \sbraket{\rho}{\sigma_{i_2}}\\
&=& \sum_{i_2} ( \sum_{i_1} c_{i_1i_2} \sket{\sigma_{i_1}} )( \sum_{{i_1}'} \bar{c}_{{i_1}'i_2} \sbra{\sigma_{{i_1}'}} )\\
&=& \sum_{i_1,{i_1}'} ( \sum_{i_2}c_{i_1i_2}\bar{c}_{{i_1}'i_2}) \sket{\sigma_{i_1}}\!\bra{\sigma_{{i_1}'}}\\ 
&=& \sum_{i,j} d_{ij} \sket{\sigma_{i}}\!\bra{\sigma_{j}}.
\eea
Similarly, if $\mathcal{L}$ is a superoperator, then
\be
\mathcal{L}\sso = \sum_{ijkl} (\mathcal{L}\sso)_{ijkl} \sket{\sigma_i\sigma_j}\!\bra{\sigma_{k}\sigma_l},
\ee
where
\bea
(\mathcal{L}\sso)_{ijkl} = \!\bra{\sigma_i\sigma_j} \mathcal{L}\sso\sket{\sigma_k\sigma_l} = \tr(\sigma_i^\dagger\sigma_j^{\dagger}\mathcal{L}[\sigma_k\sigma_l]).
\eea
\bibitem{intuition} Intuition on the advantages of this approach can be gained by considering the following simple example, involving a maximally mixed state, $\rho_I \sim I^{[L]}\otimes I^{[R]}$. Because all pure states of $\mathbb{H}^{[L]}$ are equally represented in the reduced density matrix $\rho^{[L]} \sim I^{[L]}$,  no significant truncation of this space is possible without introducing an important error. Instead, $\sket{\rho_I}$ has a one-dimensional reduced superoperator $\sSO^{[L]} \sim \sproj{I^{[L]}}$, allowing for an extreme decimation of $\mathbb{K}^{[L]}$,
\be
\left. \begin{array}{ll}
\rho \!\sim \!I^{[L]}\otimes I^{[R]}~~~\\
\!\mbox{rank}(\rho^{[L]}) = \dim(\mathbb{H}^{[L]})
\end{array}
\right\}\rightarrow \left\{
\begin{array}{ll} 
\sket{\rho} \!\sim \!\sket{I^{[L]}}\otimes \sket{I^{[R]}}\\
\mbox{rank}(\sSO^{[L]}) = 1.
\end{array}
\right.
\ee
\bibitem{compare} Given a bipartite pure state $\ket{\Psi^{[LR]}}$ with $\ch$ Schmidt coefficients $\lambda_{\alpha}$, the decomposition (\ref{eq:Schmidt}) for superket $\sket{\Psi^{[LR]}}$ will have $\ch\sso = \ch^2$ coefficients given by $\lambda\sso_{[\alpha\alpha']} = \lambda_{\alpha}\lambda_{\alpha'}$.
\bibitem{real} Coefficients $c_{i_1 \cdots i_n}$ in Eq. (\ref{eq:local}) can be made real by choosing $\sket{i_l}$ to be the superket of an Hermitian matrix in $\mathbb{L}(\mathbb{H}^{[l]})$. However, in some occasions (e.g., when computing two-point time-dependent expectation values in a state out of equilibrium) the use of complex coefficients seems unavoidable.
\bibitem{nonunitary} The evolution resulting from (\ref{eq:master2}) is in general non-unitary in the space $(\mathbb{C}_{d^2})^{\otimes n}$ of superkets $\sket{\rho}$. As explained in \cite{preparation}, the TEBD algorithm can still be used.
\bibitem{advise} It is advisable to renormalize the state $\sket{e^{-\beta H}}$ as $\sket{e^{-\beta H}/Z(\beta)}$ during the evolution by requiring $c_{0\cdots 0} = 1$ in Eqs. (\ref{eq:local})-(\ref{eq:superdeco}) at all intermediate values of $\beta$.
\end{thebibliography}
\end{document}